\documentclass[12pt,a4paper]{article}

\usepackage[utf8]{inputenc}
\usepackage[T1]{fontenc}
\usepackage{amsmath,amssymb,amsfonts}
\usepackage{graphicx}
\usepackage{xcolor}
\usepackage{booktabs}
\usepackage{array}
\usepackage[expansion=false]{microtype}
\usepackage{hyperref}
\usepackage{url}
\usepackage[
  style=authoryear,
  natbib=true,
  backend=biber,
  giveninits=true,
  uniquename=false,
  uniquelist=false,
  maxbibnames=99,
  maxcitenames=2,
  mincitenames=1,
  sortcites=true,
  isbn=false,
  url=false,
  eprint=false,
  doi=true
]{biblatex}
\AtEveryBibitem{\clearfield{month}\clearfield{day}}
\usepackage{geometry}
\usepackage{authblk}
\usepackage{float}
\usepackage{chngcntr}

\hypersetup{
  colorlinks=true,
  linkcolor=blue,
  citecolor=blue,
  urlcolor=blue
}

\newcommand{\aifssubs}{AIFS-SUBS}
\newcommand{\aifsgaia}{AIFS-SUBS-ERA5}
\newcommand{\aifscrps}{AIFS-CRPS}
\newcommand{\aifsens}{AIFS-ENS-v2}

\title{\textbf{\aifssubs{}: Extending Data-Driven Forecasting to Sub-Seasonal Timescales}}

\author[1]{Jakob Schloer}
\author[1]{Steffen Tietsche}
\author[1]{Christopher D. Roberts}
\author[1]{Lorenzo Zampieri}
\author[1]{Simon Lang}
\author[1]{Gert Mertes}
\author[1]{Gareth Jones}
\author[1]{Matthew Chantry}
\author[1]{Frederic Vitart}
\affil[1]{European Centre for Medium-Range Weather Forecasts (ECMWF)}

\begin{document}

\maketitle

\begin{abstract}
Data-driven models now rival numerical weather prediction in the medium range,
but extending them to sub-seasonal lead times raises challenges absent at
shorter horizons. Errors accumulate over long autoregressive rollouts,
systematic biases grow with lead time, and several years of data must be held
out for independent verification, even though machine-learning models otherwise
benefit from longer training records. 

To address these challenges, we adapt ECMWF's \aifscrps{} medium-range model.
\aifssubs{} adopts a 24\,h autoregressive time step to reduce error accumulation,
adds stratospheric levels and top-of-atmosphere thermal radiation as predictors,
and reserves 2007--2011 as an independent verification window. We evaluate two
configurations: \aifssubs{}, fine-tuned on operational analyses, and
\aifsgaia{}, trained on ERA5 alone. Across weeks\,2--6, \aifssubs{} matches the
operational Integrated Forecasting System (IFS) in probabilistic skill while
reducing systematic biases. For the convective (OLR) component of the
Madden--Julian Oscillation (MJO), \aifssubs{} extends skilful forecasts
(correlation\,$>$\,0.5) by eight days relative to the IFS, while matching or
exceeding the IFS for the full multivariate RMM index. \aifssubs{} also
reproduces the observed MJO modulation of tropical cyclone activity comparably.
Stratospheric skill is particularly strong with \aifssubs{}
reproducing sudden stratospheric warming (SSW) frequency and surface impact. In
the AI Weather Quest, \aifsgaia{} attains a
variable-averaged ranked probability skill score slightly ahead of the IFS at
weeks\,3 and 4. At inference,
\aifssubs{} uses about 200 times less energy than the IFS, opening the door to
much larger real-time ensembles. \aifssubs{} is ECMWF's first machine-learning
model targeted at sub-seasonal time-scales.
\end{abstract}


\section{Introduction}
\label{sec:intro}

Skilful forecasts at the sub-seasonal timescale are important for
decision-makers in agriculture, water management, energy, and disaster
preparedness, providing valuable lead time to act on upcoming weather extremes.
We define the sub-seasonal timescale as forecast lead times from week~2 to
week~6, roughly days~8--42. This range sits between the medium-range, up to
about 14\,days, where atmospheric initial conditions dominate predictability,
and the seasonal range, where slowly varying boundary conditions become the
primary source of skill. Because neither source is strong in this window, it is
often referred to as the ``predictability desert'', reflecting the difficulty of
issuing skilful forecasts at these lead times
\citep{mariotti2020,robertson2020}.

Only a handful of studies have explored global deep-learning weather prediction
(DLWP) at sub-seasonal timescales. \citet{weynSubSeasonalForecastingLarge2021}
were among the first, demonstrating that a convolutional neural network (CNN)
based ensemble can retain skill beyond two weeks but still falls short of
dynamical models. More recently, \citet{lang2026} showed that a medium-range
DLWP model retains substantial skill up to week~4 with reduced biases and higher
anomaly skill than the Integrated Forecasting System (IFS). Building tailored
systems for longer horizons, \citet{chen2024} introduced FuXi-S2S, which extends
forecasts to 42-day lead times, while \citet{ling2024} proposed FengWu-W2S,
which couples ocean--atmosphere--land states seamlessly from weather to
sub-seasonal timescales. Beyond these full global machine-learning (ML) models,
a broad range of domain-specific approaches has been developed, for instance
models targeting individual climate phenomena such as the El Ni\~no--Southern
Oscillation (ENSO) or the Madden--Julian Oscillation (MJO)
\citep[e.g.,][]{ham2019,schlor2024,delaunay2022}, alongside post-processing
models that refine dynamical or ML forecasts
\citep[e.g.,][]{bouallegue2024,guan2026,worsnop2024,roberts2026}.  In this work
we focus on global ML models, and we refer the reader to the AI Weather Quest
for a broader intercomparison of data-driven sub-seasonal forecasting
\citep{loegel2025}.

Domain-specific models can optimise a loss defined directly on their target,
such as quintile probabilities at week~4. DLWP models, by contrast, are
autoregressive and are typically trained to predict only the next global weather
state, or a handful of steps ahead. Backpropagating gradients through long
rollouts is computationally costly and numerically unstable, as gradients tend
to vanish or explode over many autoregressive steps. Extending ML models to
sub-seasonal lead times thereby introduces two specific challenges: (i) The loss
of initial-condition information in the atmosphere decays within roughly two to
three weeks \citep[e.g.,][]{judt2020}. Beyond this point the signal-to-noise
ratio drops sharply, making it difficult for a model trained on short-range
targets to learn the weak but potentially predictable signal at week~3--6 lead
times. (ii) The residual predictability is instead governed by large-scale,
slowly evolving patterns and is therefore strongly flow-dependent. At
sub-seasonal ranges, useful skill comes from phenomena such as the MJO
\citep[][and references therein]{madden1972,madden1994,zhang2005}, sudden
stratospheric warmings (SSWs) \citep[e.g.][]{baldwin2001,karpechko2017},
slowly-evolving sea surface temperature patterns, and land-surface memory
\citep[e.g.][]{koster2010}. These modes also modulate
high-impact weather, for instance the MJO's well-documented control on tropical
cyclone activity \citep[e.g.][]{camargo2009,vitart2009}, so that skill at these
lead times can translate into useful forecasts of impacts. These features open
intermittent and spatially heterogeneous windows of opportunity for skilful
forecasts \citep{mariotti2020}, and because they occur infrequently in the
training record they yield few samples from which to learn in the short
observational record.

We address these challenges by building upon the \aifscrps{} \citep{lang2026}, a
model trained to minimise the almost fair continuous ranked probability score
(afCRPS), a proper score that accounts for finite ensemble size effects such
that forcast members are rewarded when they appear to be sampled from the same
distribution as observerations (REFS). In constrast to approaches that minimize
mean squared error (MSE), this probabilistic training objective rewards ensemble
forecasts that are both sharp and statistically reliable
\citep[e.g.][]{gneiting2007}, which is crucial when forecast uncertainty is
large and flow-dependent. Our modifications to the medium-range system are as
follows. We adopt a 24\,h autoregressive time step, compared to 6\,h in the
medium-range model, to mitigate error accumulation over long rollouts. We
include stratospheric variables so that the model can represent
stratosphere--troposphere coupling and the SSW events that are a key source of
sub-seasonal predictability. We add top-of-atmosphere thermal radiation
(\texttt{ttr}), from which outgoing longwave radiation is derived, to better
diagnose the convective component of the MJO. Finally, we leave out the
years 2007 to 2011 from the training, providing an independent five-year
verification period. Unlike medium-range models, whose skill can be assessed
from many quasi-independent forecasts within a single year, sub-seasonal skill
hinges on rare events such as SSWs and MJO episodes that recur only a handful of
times per year. A multi-year window is therefore required to sample enough of
these events for statistically meaningful skill estimates, at the cost of
reducing the amount of training data.

Here, we present two model configurations. \aifssubs{} is designed for real-time
operational use and fine-tuned on operational ECMWF analyses; because this
fine-tuning ties its skill to the initialisation source. \aifsgaia{} is a
companion model trained exclusively on ERA5 reanalysis, which removes this
dependence and which we submit to the AI Weather Quest competition under the name
AIFSgaia.

\section{Methods}
\label{sec:methods}

\subsection{Data}
\label{sec:data}

\begin{figure}[!t]
  \centering
  \includegraphics[width=\linewidth]{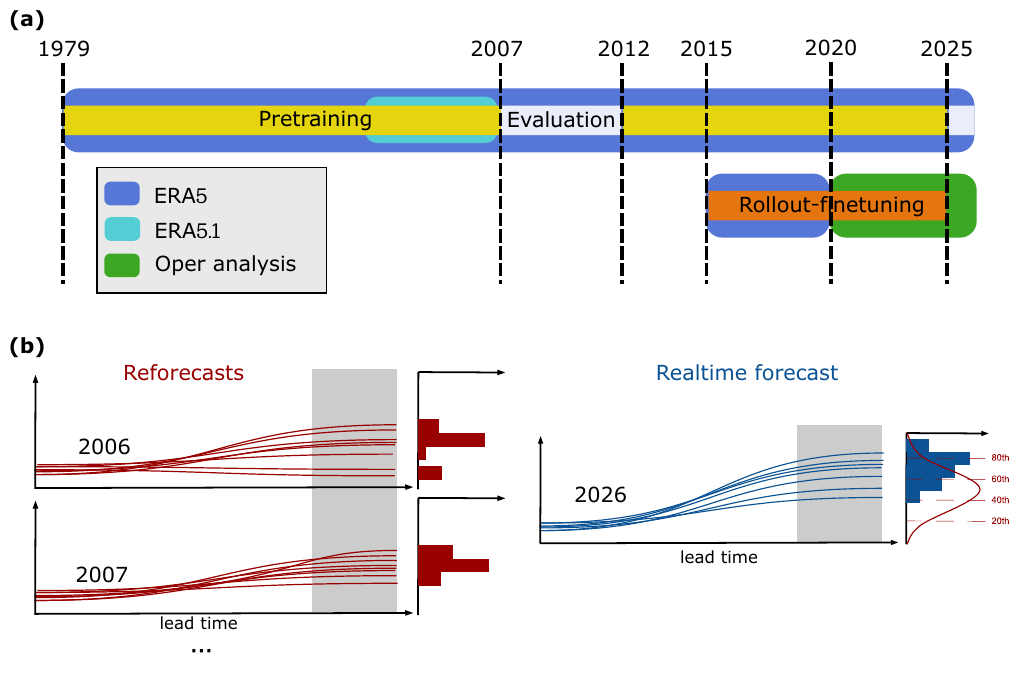}
  \caption{Training and evaluation protocol of \aifssubs{} \emph{(a)}. We first
  train for a single-step 24\,h forecast on ERA5 and then fine-tuned with a
  3-step autoregressive rollout on a combined dataset of ERA5 and operational
  analyses. For evaluation, we withhold 2007--2011 as an independent
  verification period.   For each target date, a forecast ensemble is generated
  alongside a set of reforecasts initialised on the same month--day in each of
  the 20 preceding years (\emph{b}). The resulting reforecast distribution
  defines the model climatology against which the forecast members are
  calibrated, removing systematic biases.}
  \label{fig:datasplit}
\end{figure}

All models are trained on ERA5 \citep{hersbach2020}, the fifth-generation ECMWF
atmospheric reanalysis, at 6\,h temporal resolution on an O96 grid
(approximately 0.9$^\circ$). The standard ERA5 dataset contains a known warm
bias in the lower stratosphere for the period 2000--2006. We correct for this bias by
using the upper atmospheric levels ($\leq$ 300 hPa) from ERA5.1
\citep{simmons2020}, which provides an improved stratospheric analysis
for this period and ensures consistent upper-level boundary conditions
throughout training.

Since \aifssubs{} is designed for realtime use and will be initialised from
operational ECMWF analyses, we include such analyses in the fine-tuning phase.
ERA5 and the operational analysis differ both in model configuration and in
their treatment of the lower boundary. ERA5 was produced with IFS Cycle 41r2 at
a horizontal resolution of approximately 31\,km, whereas the current operational
analysis uses IFS Cycle 50r1 at approximately 9\,km resolution. ERA5 uses
externally prescribed daily sea-surface temperature and sea-ice boundary
conditions, whereas in Cycle 50r1 the ocean and sea-ice states are updated within
the outer loops of the atmospheric 4D-Var analysis. Such differences can lead to
a distribution shift when a model trained or fine-tuned on one analysis product
is initialised from the other. 

Table~\ref{tab:variables} lists all input and output variables. Compared to the
medium-range \aifscrps{}, the key additions are top-of-atmosphere thermal
radiation (\texttt{ttr}) and a set of stratospheric pressure levels reaching up
to 2\,hPa.

\begin{table}[!b]
  \centering
  \caption{Variables used by \aifssubs. Prognostic variables are both input and
  output, diagnostic are output, and forcing variables are input only.}
  \label{tab:variables}
  \begin{tabular}{l|llr}
    \toprule
    \textbf{Category} & \textbf{Variables} & \textbf{Levels}  & \textbf{Type} \\
    \midrule
    Upper-air & Geopotential ($z$) & 2, 5, 10, 30, 50, 70, 100, 150, & prognostic \\
              & Temperature ($t$) &  200, 250, 300, 400, 500, 600, & \\
              & U/V wind & 700, 850, 925, 1000 & \\
    \cmidrule(l){2-3}
              & Specific humidity ($q$) & same levels but only $>=$ 150\,hPa & \\
    \cmidrule(l){2-4}
              & W wind &  same levels & diagnostic \\
    \midrule
    Single-level & \multicolumn{2}{c}{10 metre u wind, 10 metre v wind component,} & prognostic \\
                 & \multicolumn{2}{c}{2 metre dewpoint temperature, 2 metre temperature,} & \\
                 & \multicolumn{2}{c}{Mean sea level pressure, Surface pressure,} & \\
                 & \multicolumn{2}{c}{Total column water, Low cloud cover,} & \\
                 & \multicolumn{2}{c}{Medium cloud cover, Total cloud cover,} & \\
                 & \multicolumn{2}{c}{Skin temperature} & \\
    \cmidrule(l){2-4}
                 & \multicolumn{2}{c}{Total precipitation, Convective precipitation,} & diagnostic \\
                 & \multicolumn{2}{c}{Snowfall, Total cloud cover,} & \\
                 & \multicolumn{2}{c}{Low cloud cover, Medium cloud cover,} & \\
                 & \multicolumn{2}{c}{Top net thermal radiation, Surface solar radiation downwards,} & \\
                 & \multicolumn{2}{c}{Surface thermal radiation downwards} & \\
    \cmidrule(l){2-4}
                 & \multicolumn{2}{c}{Insolation, Land--sea mask, orography,} & forcing \\
                 & \multicolumn{2}{c}{Std.\ dev.\ of orography, Slope of sub-gridscale orography} & \\
    \bottomrule
  \end{tabular}
\end{table}

\subsection{Model Architecture}
\label{sec:model}

The \aifssubs{} architecture follows the encoder--processor--decoder design of
\aifscrps{} \citep{lang2026}, which builds on the AIFS graph neural network
framework \citep{langAIFSECMWFDatadriven2024}. It combines a graph-transformer
neural network encoder and decoder with a sliding-window transformer processor.
The encoder maps gridded input fields onto an O48 latent grid using a
graph-attention layer. Attention in the processor is computed along spiral
longitudinal bands \citep{langAIFSECMWFDatadriven2024}. The processor comprises
16 layers with an embedding dimension of 1024 and 8 attention heads, totalling
approximately 230 million trainable parameters. Forecast ensembles are generated
by conditioning on random noise: for each member $i$, independent samples $z_i$
drawn from a standard normal distribution (dimension = latent grid $\times$ 4
noise channels) are transformed by a two-layer MLP and injected via conditional
layer normalisation in each processor layer.

Given the atmospheric state $x$ at times $t-24$\,h and $t$, each ensemble member
$i$ predicts the 24-hour tendency, such that the next state is obtained as
\begin{equation}
  \hat{x}_i(t + 24\,\mathrm{h}) \;=\; x(t) \;+\;
  f_{\Theta}\!\left( x(t),\; x(t - 24\,\mathrm{h}),\; z_i \right),
  \label{eq:autoregressive}
\end{equation}
where $f_{\Theta}$ denotes the encoder--processor--decoder network with
parameters $\Theta$ and $z_i$ the member-specific noise sample described above.
Forecasts at longer leads are produced by iterating Eq.~\ref{eq:autoregressive}
autoregressively. This longer time step --- compared to 6\,h in the medium-range
model --- reduces the number of rollout steps required to reach week~6 from
${\sim}240$ to ${\sim}42$, substantially decreasing both error accumulation and
inference cost. On ECMWF's supercomputer, a single-member 46-day forecast of
\aifssubs{} uses about 200 times less energy than the corresponding IFS run. We
report energy because the IFS runs on CPUs while \aifssubs{} runs on GPUs. In
terms of compute time producing a forecast, \aifssubs{} is roughly 920 times
faster. However, we note that the two models differ considerably in resolution
(O320, for the IFS versus O96 for \aifssubs{}) and number of variables (137
vertical levels for IFS versus 18 for \aifssubs{}).

\subsection{Training}
\label{sec:training}

Training proceeds in two stages: pre-training on single-step targets followed
by fine-tuning with multi-step autoregressive rollouts. Throughout both stages,
we maintain Exponential Moving Average (EMA) weights of the model parameters
\citep{polyak1992,tarvainen2017}. EMA smooths parameter updates across training
iterations and proves important for the stability of long autoregressive
rollouts at inference time. 

\paragraph{Pre-training.}
Both \aifssubs{} and \aifsgaia{} are pre-trained on ERA5 for the period
1979--2006 and 2012--2024, excluding the held-out 2007--2011 verification
window, see Fig.~\ref{fig:datasplit} (a). The objective is a single-step 24\,h
forecast, allowing the model to learn accurate short-range dynamics before being
exposed to the multi-step regime. We train with ensemble size 4, for 300k iterations with an effective
batch size of 16 on 16 NVIDIA A100 GPUs (${\sim}$4 days wall-clock time),
using AdamW with cosine annealing (peak learning rate $10^{-3}$, 1000 warm-up
iterations).

\paragraph{Fine-tuning \aifsgaia{}.}
\aifsgaia{} is fine-tuned with a 3-step (3-day) autoregressive rollout on ERA5
for 2015--2024, keeping the training signal entirely within the reanalysis. This
configuration provides a clean ERA5-only baseline and serves as the basis for
participation in the AI Weather Quest competition. Fine-tuning runs for 50k
iterations with a reduced peak learning rate of $5 \times 10^{-5}$, retaining
the same batch size and hardware setup. We selected the 3-step rollout
by ablation; extending the rollout beyond three steps did not improve forecast
scores, so we retain three steps to limit fine-tuning cost.

\paragraph{Fine-tuning \aifssubs{}.}
\aifssubs{} is fine-tuned with a 3-step rollout on a combined dataset: ERA5 for
2015--2019 and operational ECMWF analyses for 2020--2024. The combined dataset
is then randomly shuffled, drawing minibatches across both eras. The operational
analyses span multiple IFS versions, introducing version-to-version variability
that encourages generalisation to real-time forecasting conditions. The
optimisation schedule mirrors that of \aifsgaia{}.

\subsection{Evaluation of sub-seasonal forecasts}
\label{sec:eval}

Sub-seasonal evaluation differs from medium-range evaluation in three aspects:
we score (i) weekly averages rather than instantaneous fields, because
day-to-day variability is large relative to the predictable signal; (ii)
anomalies relative to each model's own climatology, which removes systematic
biases that would otherwise dominate, and (iii) indices derived by projecting
anomalies onto pre-defined patterns to provide low-dimensional evaluation of
large-scale circulation (e.g. MJO, polar vortex index).

Designing an evaluation protocol for ML-based sub-seasonal forecasting involves
several constraints compared to dynamical models. First, robust skill estimates
require a sufficiently long verification period, yet any year withheld for
evaluation is a year unavailable for training potentially leading to lower
forecast skill.  Second, climate change introduces a distribution shift, which
favours training on the most recent data so that the model operates in an
approximately stationary regime. Third, the quality of reanalysis used for
training is not uniform in time and degrades noticeably in the pre-satellite
era. Fourth, \aifssubs{} is fine-tuned on operational analyses, which restricts
that part of the training signal to the most recent years.

Balancing these constraints, we withhold 2007--2011 from all training stages
and use it as an independent five-year verification set. Five years lies at
the lower end of what the S2S community typically uses (around twenty years),
but it preserves the bulk of the recent record for training while still
providing a meaningful number of independent forecast cases. We chose this
particular window because it spans several phenomena that are central to
sub-seasonal predictability:
\begin{itemize}
  \item ENSO: the 2009--10 El Ni\~no and the strong 2010--11 La Ni\~na.
  \item Sudden stratospheric warmings: multiple major events, including
  February 2008, January 2009, and February/March 2010.
  \item Modern satellite coverage, which ensures consistent data quality and a
  realistic test of the model's ability to real-time forecasting.
\end{itemize}

We complement the reforecast evaluation with forecasts submitted to the AI
Weather Quest competition over the period mid-August 2025 to mid-February
2026. Although limited to six months of weekly forecasts, these test the model on
future cases and thus provide an independent estimate
of real-world performance. Throughout, ERA5 is the verification reference (ground
truth) for all variables, except the tropical-cyclone analysis, which verifies
against IBTrACS observations. The full evaluation protocol is illustrated in
Fig.~\ref{fig:datasplit}(b) and detailed in Sec.~\ref{sec:si_evaluation} of the
Supplementary Material.

\section{Results}
\label{sec:results}
\subsection{Global biases and skill}

\begin{figure}[!b]
  \centering
  \includegraphics[width=\linewidth]{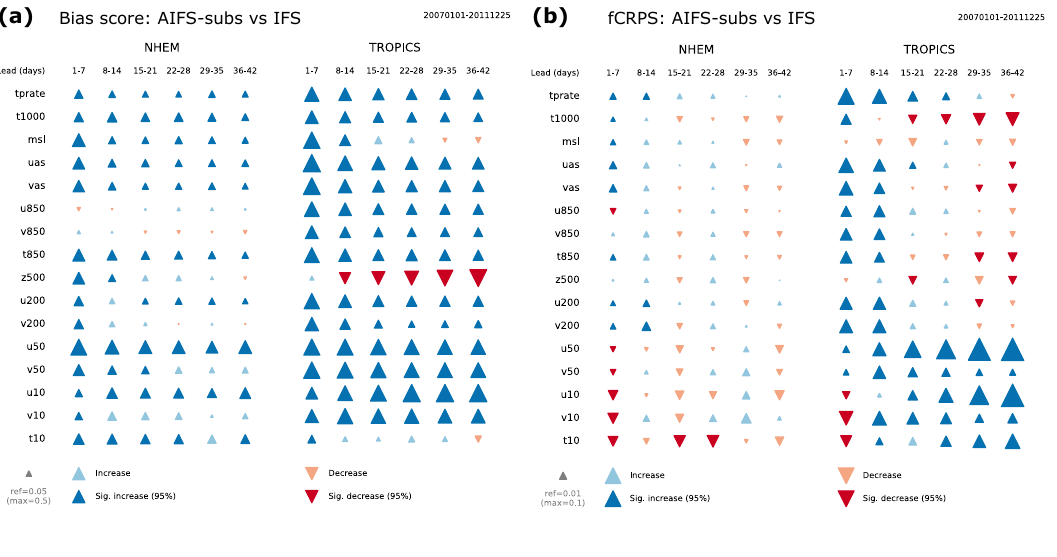}
  \caption{
    \aifssubs{} significantly reduces biases (\emph{a}) relative to IFS (49r1) for
    key surface and upper-air variables at week\,1 through week\,6 averaged over
    the Northern Hemisphere and Tropics. For the probabilistic fCRPSS computed
    over anomalies (\emph{b}), \aifssubs{} shows significant improvement in week 1
    and 2 for most variables and neutral scores for week 3 and beyond in the
    Northern Hemisphere. In the Tropics, surface and tropospheric variables improve
    significantly in weeks 1 and 2, while stratospheric variables improve across
    all lead times. Scores are computed over the 5-year verification period
    2007-2011 with 5 initialisation days per month.
    }
  \label{fig:scorecard}
\end{figure}

We first verify \aifssubs{} in a controlled reforecast setting, which yields
statistically robust results across the full set of predicted variables. We use
the five-year period 2007--2011, which is held out from training, with five
initialisation dates per month (the 1st, 7th, 13th, 19th, and 25th) and ten
ensemble members.

We follow \citet{roberts2026a} and separately evaluate climatological mean state
and forecast anomalies calculated following the ‘by-member–other-years’ method
described in \citet{roberts2025}, which references each forecast member to a
climatology built from the remaining verification years and stays unbiased for
short reforecast periods. Impacts on the mean state are assessed by computing
relative changes in the mean absolute bias (MAB) using the mean absolute bias
score (MABS; Eq.~\ref{eq:mabs} in appendix Sec.~\ref{sec:si_mabs}).
Probabilistic skill of anomaly forecasts is evaluated using the fair version of
the continuous ranked probability score
\citep[fCRPS,][]{ferro2008,ferro2014,leutbecher2019}. Skill scores are computed
from area-weighted scores averaged over all forecast cases for a given lead
time, following Eqs.~\ref{eq:fcrps}--\ref{eq:fcrpss} in appendix
Sec.~\ref{sec:si_fcrps}.

\aifssubs{} significantly reduces biases relative to the IFS (49r1) for 
surface and upper-air variables from week\,1 through week\,6
(Fig.~\ref{fig:scorecard}a). In the Northern Hemisphere, the reduction is
variable and lead-time dependent: most variables and lead times improve
significantly, while a few, such as u850 and v850, retain biases comparable to
the IFS. The bias reduction is strongest in the Tropics ($>5\%$), with two
exceptions: 500\,hPa geopotential height, where \aifssubs{} is worse than the
IFS, and mean sea-level pressure beyond week\,3, where the two are comparable.

As in the real-time evaluation, we focus on anomaly scores relative to the model
climatology, computing the anomalies of each year with respect to the
climatology over the remaining four years in the verification period. The fCRPSS
between \aifssubs{} and the IFS gives a mixed picture
(Fig.~\ref{fig:scorecard}b). In the Northern Hemisphere, \aifssubs{} is on par
with the IFS for most variables and lead times; surface variables in week\,1
improve slightly but significantly, whereas stratospheric variables in week\,1
are slightly degraded. In the Tropics, \aifssubs{} improves significantly over
the IFS in weeks\,1 and 2 for surface and tropospheric variables, and is on par
thereafter, although 2\,m temperature, t850, and z500 are notably worse in
weeks\,5 and 6. The stratospheric variables, by contrast, improve significantly
across all lead times in the Tropics, except for u, v, and t at 10\,hPa in week
1. The improvement is largest for zonal wind at 50\,hPa, suggesting a better
representation of the quasi-biennial oscillation (QBO) in \aifssubs{}.

\subsection{Madden--Julian Oscillation}

\begin{figure}[!t]
  \centering
  \includegraphics[width=\linewidth]{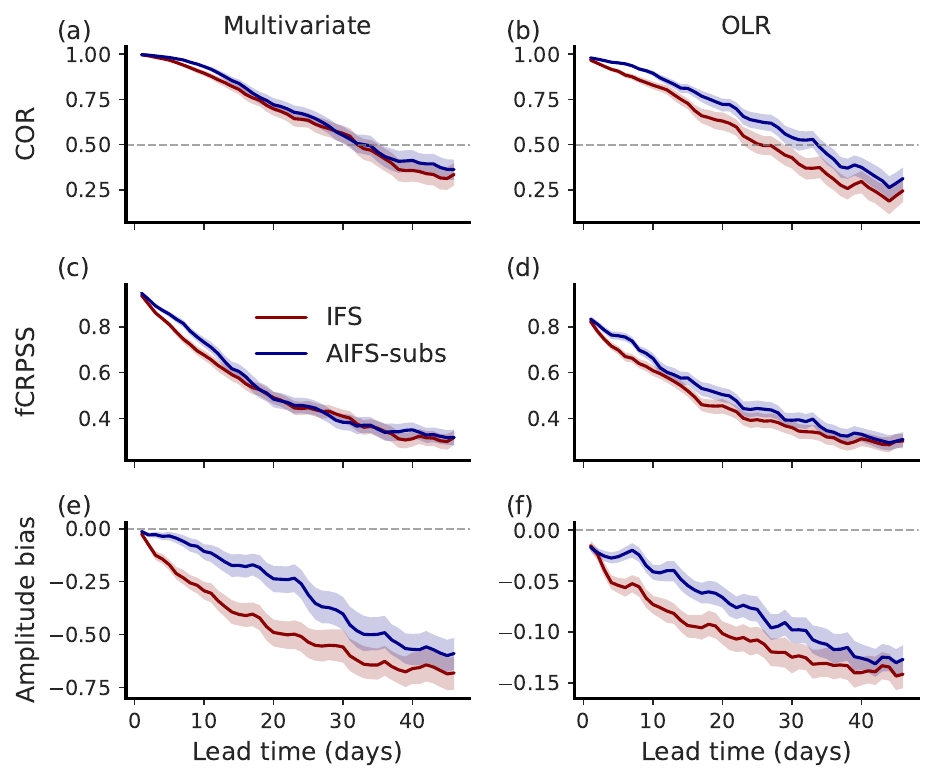}
  \caption{MJO forecast skill of \aifssubs{} and IFS as a function of forecast
  lead time over the 2007–2011 verification period. Bivariate correlation (COR)
  \emph{(a, b)}, fCRPSS \emph{(c, d)}, and amplitude bias \emph{(e, f)} are computed from
  the bivariate RMM index derived from OLR/U200/U850 (left column) and its OLR
  component, i.e. the OLR anomaly projected onto the OLR part of the RMM EOFs
  (right column).
  The dashed line in \emph{(a, b)} marks the COR = 0.5 skill threshold. Shadings denote the
  95\% confidence interval from 500 bootstrap resamples of forecast start
  dates.}
  \label{fig:mjo}
\end{figure}

To evaluate the MJO we use the bivariate Real-Time Multivariate MJO (RMM) index
\citep{wheeler2004,gottschalck2010} calculated for IFS and AIFS
forecasts as described in \citet{roberts2026b} and references therein. In
brief, the two components of the bivariate index (RMM1 and RMM2) are derived by
projecting daily mean anomalies onto the two leading observation-based
multivariate EOFs of meridionally averaged (15$^{\circ}$S--15$^{\circ}$N) zonal
winds at 850\,hPa and 200\,hPa and outgoing longwave radiation (OLR). MJO
amplitude and phase are defined as $\sqrt{\textnormal{RMM1}^2 +
\textnormal{RMM2}^2}$ and $\textnormal{arctan2}(\textnormal{RMM2},
\textnormal{RMM1})$, respectively. Phase numbers correspond to the different
sectors of the MJO phase diagram and are indicative of MJO activity over the
Indian Ocean (phases 2 and 3), maritime continent (phases 4 and 5), western
Pacific Ocean (phases 6 and 7), and the Atlantic Ocean/Africa (phases 8 and 1).
Unlike previous versions of \aifscrps{} tested at sub-seasonal lead times,
\aifssubs{} includes top-of-atmosphere thermal radiation (\texttt{ttr}) as a
diagnostic variable, from which we derive outgoing longwave radiation (OLR),
which allows us to separately evaluate the convective component of the MJO RMM
index.

Fig.~\ref{fig:mjo} shows the MJO forecast skill of \aifssubs{} and the IFS as a
function of lead time over the held-out verification period. We evaluate two
forms of the RMM index: the full multivariate RMM index derived from
OLR/U200/U850 (left column) and its OLR component alone (right column).

For the full multivariate RMM index, \aifssubs{} either matches or exceeds IFS
performance at all lead times, both in the bivariate correlation
(Fig.~\ref{fig:mjo}a) and in the probabilistic fCRPSS, where the difference is
not significant beyond day~12 (Fig.~\ref{fig:mjo}c). The improvement from
\aifssubs{} is strongest in the OLR component of the index. Using a correlation
threshold of 0.5 for a skilful forecast, 10-member forecasts with \aifssubs{}
retain skill in the OLR component out to 33 days, compared with 25 days for the
IFS, a statistically significant gain of eight days (Fig.~\ref{fig:mjo}b), with a
matching significant improvement in fCRPSS (Fig.~\ref{fig:mjo}d). This indicates
that the main improvements to the multivariate RMM index are coming from
improvements to the representation of tropical convection rather than the
tropical wind field that dominates the full RMM index.

Both models damp the MJO with increasing lead time and exhibit an amplitude
bias, a common shortcoming of sub-seasonal forecasts. For \aifssubs{}, however,
this bias is significantly smaller than for the IFS across all lead times and
for both index definitions (Fig.~\ref{fig:mjo}e,~f), indicating a better
representation of the MJO's strength. The reduced amplitude bias is particularly relevant
because dynamical models tend to underestimate MJO amplitude, a leading source
of error in their representation of MJO teleconnections \citep{vitart2017}.

Nevertheless, we caution that these skill differences are estimated over a single
five-year verification window and may be sensitive to the period of evaluation,
given the small number of independent MJO events it contains. More extensive
real-time testing across a wider range of cases will be necessary to confirm that
the gains in the OLR component of the RMM index are robust.

\begin{figure}[!t]
  \centering
  \includegraphics[width=\linewidth]{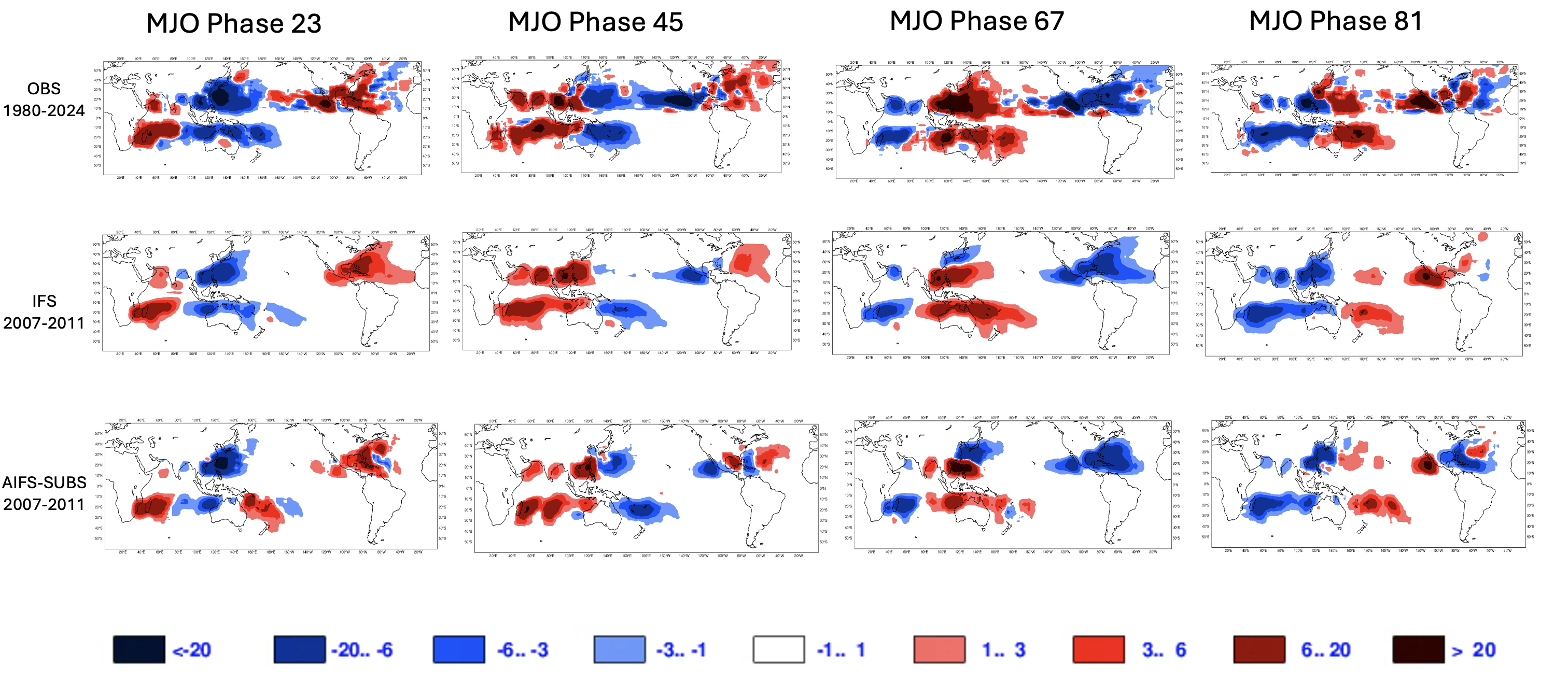}
  \caption{Anomaly in the number of tropical storms within a 300\,km radius per
  day ($\times$1000) for different MJO phases, averaged over JJASON in the
  2007--2011 verification period. A tropical storm day is counted when the wind
  speed exceeds 17\,m\,s$^{-1}$. Counts are normalised by the number of years,
  ensemble members, and daily output steps, and shown as the anomaly relative to
  the all-phase climatology. The top row shows the observed relationship from
  IBTrACS, while the middle and bottom rows show the same for IFS and
  \aifsgaia{} reforecasts, respectively.}
  \label{fig:mjo_tc_teleconnection}
\end{figure}

The MJO is a well-documented modulator of tropical cyclone (TC) activity,
shifting the regions of enhanced and suppressed genesis as its convective
envelope propagates eastward \citep[e.g.,][]{camargo2009,vitart2009}. We assess whether
\aifsgaia{} captures this teleconnection by comparing the anomaly in the number
of tropical storms per day and MJO phase in observations (IBTrACS
\citet{knapp2010}), the IFS, and \aifsgaia{} over JJASON
(Fig.~\ref{fig:mjo_tc_teleconnection}). Counts are normalised by the number of
years, ensemble members, and daily output frequency, and expressed relative to
the all-phase climatology, so that the single observed realisation and the
multi-member reforecasts are directly comparable.  Both the IFS and \aifsgaia{}
reproduce the observed modulation well, in the training period
(Fig.~\ref{fig:si_mjo_tc_teleconnection}) and the out-of-training verification
period (Fig.~\ref{fig:mjo_tc_teleconnection}) alike: TC activity increases where
the MJO enhances convection and decreases where convection is suppressed,
producing the characteristic eastward propagation of the activity anomaly.
Notably, \aifsgaia{} reproduces this relationship in the shorter out-of-training
period almost equivalently as in the 15-years withing the training period
(Fig.~\ref{fig:si_mjo_tc_teleconnection}). Two features are less well captured:
the increase in TC activity in the South Pacific during phases~2--3, and the
decrease over the northern part of the north-western Pacific during phases~6--7.
These discrepancies aside, the agreement lends confidence that \aifsgaia{} can
be used for sub-seasonal forecasts of TC activity.

\subsection{Sudden stratospheric warming events}

\begin{figure}[!b]
  \centering
  \includegraphics[width=\linewidth]{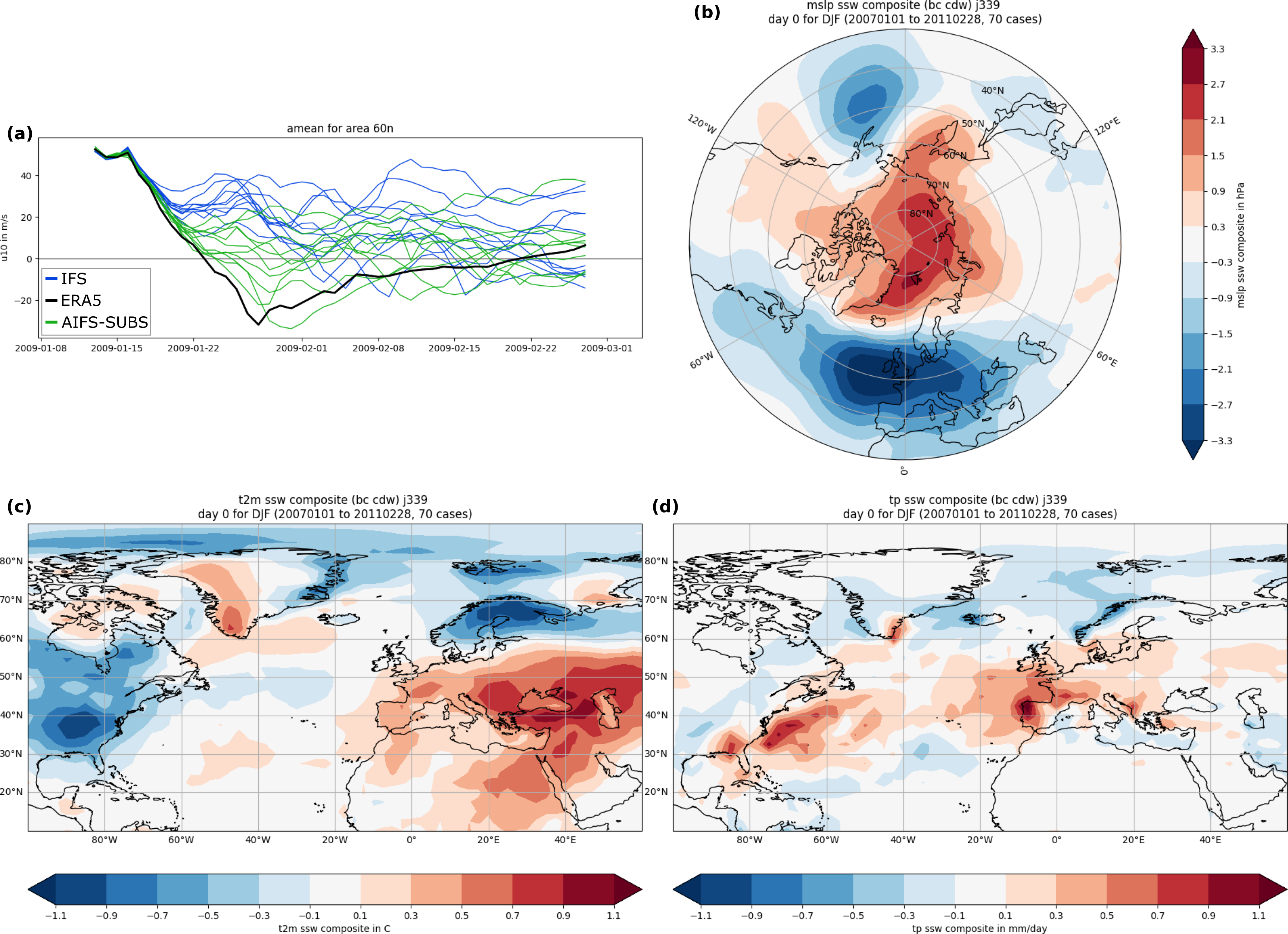}
  \caption{Forecast plumes of \aifssubs{} and IFS for the SSW event in January
  2009 with forecast initialization on 13 Jan are shown in (a).  Surface impact
  composites 10-35 days from all SSW events occurring in the 2007-2011
  reforecasts for (b) mean sea level pressure, (c) 2m temperature, and (d) total
  precipitation. See supplementary material for the equivalent composites in
  49R1 IFS reforecasts.}
  \label{fig:ssw} 
\end{figure}

The stratosphere is a known source of sub-seasonal predictability
\citep{domeisen2020}. Therefore, \aifssubs{} is trained on and predicts
additional levels in the stratosphere (see Table~\ref{tab:variables}). Training
data for the stratospheric levels is ERA5, with two modifications: First, to
avoid severe stratospheric errors present in ERA5 between 2000-2006, we train on
the corrected reanalysis ERA5.1 for these years \citep{simmons2020}. Second, we
found that training on the ERA5 stratospheric humidity is detrimental to
forecast quality, a fact that we ascribe to the limited quality of the ERA5
stratospheric humidity analysis. \aifssubs{} therefore does not include humidity
on levels from 100 hPa upwards.

When comparing \aifssubs{} with and without additional stratospheric levels in
pre-training (not shown), we find that including stratospheric levels improves
the middle and lower tropospheric forecast scores by 1-2\% for Northern
Hemisphere winter, whereas in other regions and seasons there is limited
tropospheric impact.  However, forecast scores for the upper troposphere and
lower stratosphere (200 – 50 hPa) are significantly improved.

Fig.~\ref{fig:scorecard} shows that the final fine-tuned \aifssubs{} provides
predictions that are much improved over IFS beyond week one for the lower
stratosphere (50 hPa and 10 hPa) in the tropics, whereas they are on par with
IFS in the Northern Hemisphere extratropics. Predictions of the few major
Northern Hemisphere SSWs during the 2007-2011 test period are also competitive
with IFS. As an example, Fig.~\ref{fig:ssw}a shows forecasts of the January
2009 SSW event for IFS and \aifssubs{} initialized on 13 January. For this
event, half of the members of \aifssubs{} correctly predict a disruption of the
stratospheric polar vortex, indicated by a reversal of the 60N zonal-mean zonal
winds at 10 hPa, whereas no IFS member predicts this event. 

Evolution and surface impact of SSWs varies substantially from case to case, so
the 4 events happening in the test period 2007-2011 are too small a sample to
draw robust conclusions. We therefore use the approach described in
\citet{spaeth2022} and provide statistics on all SSW that are generated
within the reforecast ensemble. We find that \aifssubs{} ensemble members
throughout all Dec-Feb 2007-2011 reforecasts generate 112 SSW within forecast
day 10 to 36. This corresponds to a frequency of about 1 SSW every two winters.
49\% of these events have strong surface impact as defined in
\citet{karpechko2017}. The corresponding numbers for the IFS 49R1
reforecasts are 98 SSW, 51\% of which have strong surface impact. Both the
frequency of generation and the fraction of surface impact are thus similar in
\aifssubs{} and IFS.  Considering the substantial event-to-event variability, this
seems consistent with the observational record.

The spatial patterns of surface impact warrant some further attention. Canonical
patterns of mean sea-level pressure, 2m temperature and precipitation have been
reported \citep[e.g.,][]{butler2017} and, especially in Europe,
are often used to anticipate winter cold spells in the sub-seasonal range.
Fig.~\ref{fig:ssw} b-d show composites of these surface parameters after SSWs generated by
\aifssubs{}. A clear negative phase of the Arctic Oscillation is evident,
alongside consistent anomalies in 2m temperature and precipitation: cold and dry
in Northern Europe, warm and wet in Southern Europe. These SSW surface impacts
are broadly consistent with composites from the observational record. Overall,
the surface impact composites in \aifssubs{} are very similar to those in IFS (see
Fig.~\ref{fig:si_ssw_ifs} in the Appendix). 

A few caveats apply to the reforecast evaluation from 2007-2011. We use ERA5
both to initialise and to verify \aifssubs{}, except for the TC analysis which
is based on observations. Because \aifssubs{} is fine-tuned on operational
analyses, we expect the reported scores to underestimate the skill of real-time
forecasts initialised from operational analyses. The comparison with the IFS
remains fair, however, as the IFS reforecasts are also verified against ERA5.
The five-year verification period is relatively short, but it is a compromise
that preserves a sufficiently long training period while retaining an
independent verification set. We also note that the 2007--2011 window is
bracketed by training data (1979--2006 and 2012--2024), so it is not a strict
temporal-causal split: unlike the live AI Weather Quest period (mid-August 2025
to mid-February 2026), which is a genuine out-of-sample future test, the model
may draw on climate and trend information from years on both sides of the
verification window.

\subsection{AI Weather Quest results}

\begin{figure}[!t]
  \centering
  \includegraphics[width=\linewidth]{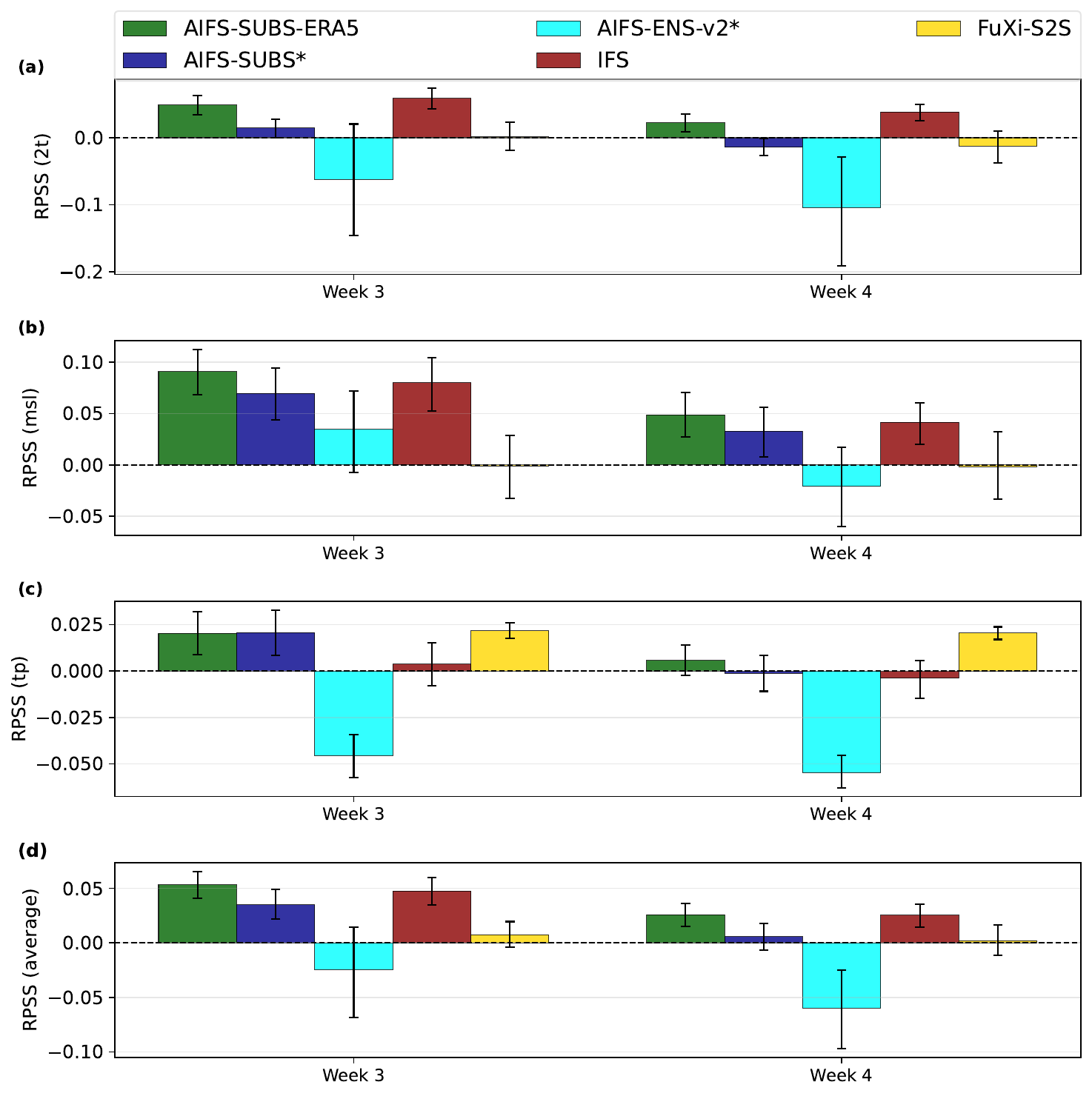}
  \caption{Period-aggregated Ranked Probability Skill Score (RPSS) for the AI
  Weather Quest evaluation (mid-August 2025 to mid-February 2026), showing
  (\emph{a}) 2\,m temperature, (\emph{b}) mean sea-level pressure, (\emph{c})
  total precipitation, and (\emph{d}) the average across the three variables of
  the quintile probability forecasts. Both \aifssubs{} and \aifsens{} were not
  submitted to the competition and are marked with an asterisk (*) because they
  are initialised here from ERA5 rather than from the operational analyses they
  were fine-tuned on; their scores therefore do not reflect the skill
  expected in operational use. Error bars indicate 90\,\% confidence intervals
  estimated via bootstrapping.}
  \label{fig:quest_rpss}
\end{figure}

The reforecast verification above is based on the ‘by-member–other-years’
anomaly calculations \citep{roberts2026a} in the period 2007-2011. We complement
it with real-time forecasts submitted to the AI Weather Quest competition, a
challenge to compare data-driven global sub-seasonal forecasts
\citep{loegel2025}. We submit \aifsgaia{} as a weekly entry to the competition,
which compares sub-seasonal forecasts from operational and experimental systems
against a common protocol. The required outputs are quintile probability
forecasts of weekly-mean 2\,m temperature (land only), mean sea-level pressure
(global), and total precipitation (land only) at week\,3 and week\,4 lead times.
Forecasts are issued weekly and the period analysed here spans 29 weeks from
mid-August 2025 to mid-February 2026.

For each target date and lead time, the model climatology is constructed from a
10-member reforecast ensemble initialised on five dates $\{-4, -2, 0, +2, +4\}$
around the corresponding month--day in each of the 20 preceding years, yielding
$N_{\mathrm{c}} = 1000$ hindcast samples per target date (see
Fig.~\ref{fig:datasplit}(b)). The real-time forecasts themselves are issued with
an ensemble of $N_{\mathrm{e}} = 200$ members. The quintile thresholds are
estimated from the climatological distribution and the target probabilities are
defined analogously from ERA5, following \citet{loegel2025}. Following the
competition protocol, skill is reported as the RPSS relative to a uniform
climatological reference, see also \ref{sec:si_evaluation} in the Appendix. 
Given the limited number of forecast cases, confidence intervals are estimated
by bootstrapping the forecast cases with replacement 1000 times and recomputing
the RPSS for each resample.

We compare \aifsgaia{} against a leading physics-based model, ECMWF's IFS, and a
ML model tailored for sub-seasonal forecasting, FuXi-S2S.

The \textbf{IFS} ensemble is a leading operational physics-based S2S
system and serves as our primary benchmark. We use the operational cycle 49r1,
whose 51-member ensemble derives its spread from initial-condition perturbations
(Ensemble of Data Assimilations and singular-vector perturbations) and
stochastic model-physics perturbations \citep{lang2012,lang2021,leutbecher2008}.

\textbf{FuXi-S2S} acts as a ML sub-seasonal baseline
\citep{chen2024}, which extends the FuXi medium-range architecture to 42-day
lead times and directly generates quintile probabilities. Rather
than rerunning the model, we use its AI Weather Quest submissions, entered as
\emph{Fengshun}, which the China Meteorological Administration (CMA) reports to
correspond to FuXi-S2S.

Further, we include \aifssubs{}* and \aifsens{}* in Fig.~\ref{fig:quest_rpss}. 
\aifsens{} is ECMWF's operational medium-range ML ensemble \citep{lang2026}, run
at $\sim$0.25\textdegree{} (N320, full Gaussian grid with 320 latitude circles)
with a 6\,h time step, pre-trained on ERA5 and fine-tuned on operational
analyses with a 12-step (72\,h) rollout. 
Both models, denoted with an asterisk, are here initialised from ERA5 for both
reforecasts and real-time forecasts, which departs from their operational setup:
in operations they are initialised from operational analyses (version 50r1),
which only became available in May 2026, after the evaluation window. While the
RPSS scores of \aifssubs{}* and \aifsens{}* are therefore not directly
comparable to the other models, they provide a useful reference for the impact
of the initialisation source and motivate the design choices of \aifssubs{}.

\begin{figure}[!t]
  \centering
  \includegraphics[width=\linewidth]{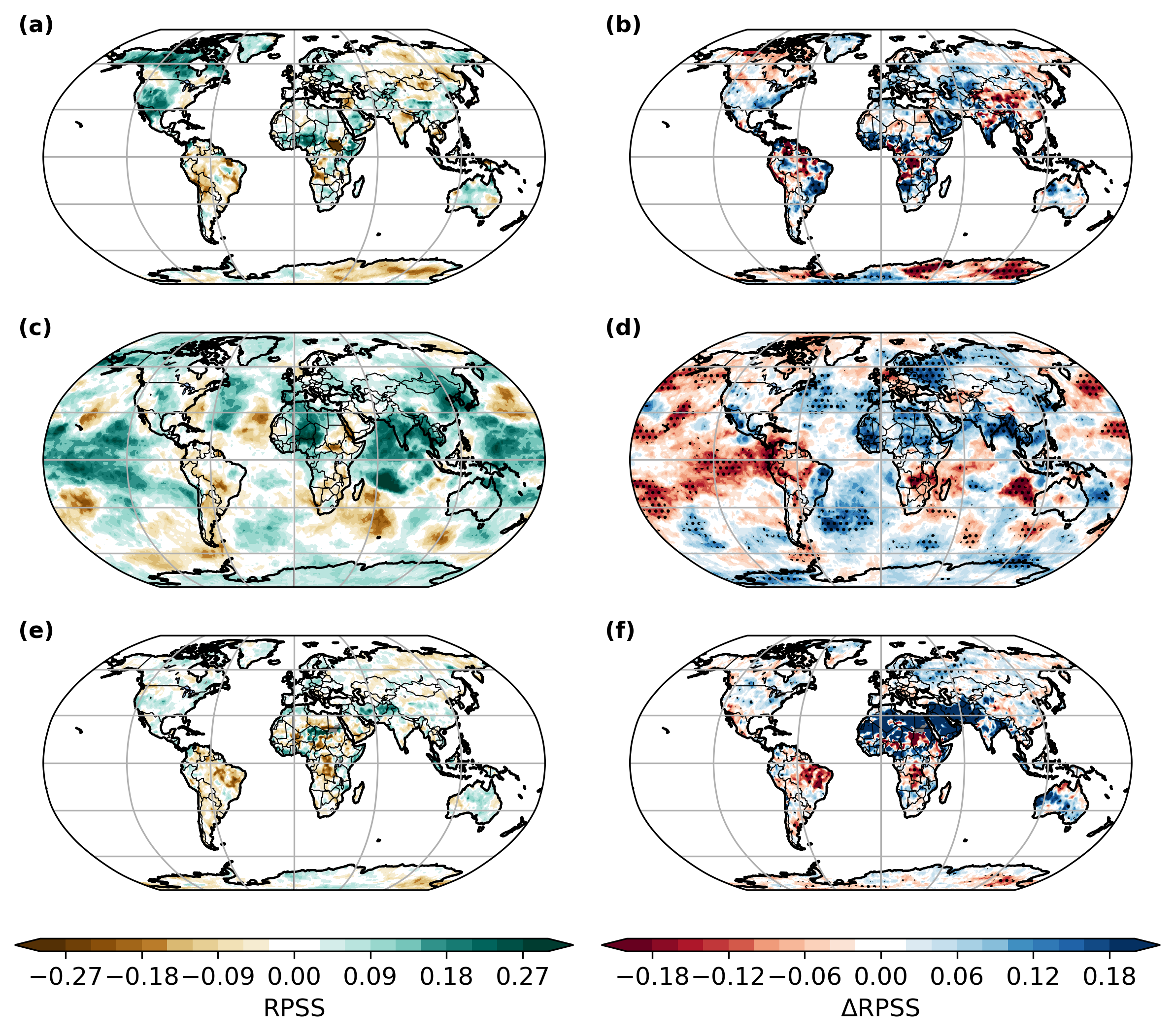}
  \caption{RPSS score maps for 2m temperature (a), mean sea level pressure (c),
  and total precipitation (e) for week~3 of \aifsgaia{} against climatology.
  Differences in RPSS relative to IFS are shown in (b, d, f) with blue
  indicating an improvement and red indicating a degradation with respect to the
  IFS. Stippling indicates grid cells where the RPSS difference is statistically
  significant at the 90\,\% confidence level, estimated via bootstrapping.}
  \label{fig:quest_maps}
\end{figure}

Period-aggregated RPSS scores are summarised in Fig.~\ref{fig:quest_rpss}.
Averaged across the three variables, \aifsgaia{} scores are comparable to the
IFS and significantly higher than the ML baseline, FuXi-S2S
(Fig.~\ref{fig:quest_rpss}d). The ranking varies by variable. For 2\,m
temperature, the IFS attains the highest RPSS at both week\,3 and week\,4,
followed by \aifsgaia{} (Fig.~\ref{fig:quest_rpss}a). For mean sea-level
pressure, \aifsgaia{} is on par with the IFS and again outperforms FuXi-S2S
(Fig.~\ref{fig:quest_rpss}b). For total precipitation, \aifsgaia{} significantly
outperforms the IFS at week\,3, while FuXi-S2S shows the highest central RPSS at
both lead times (Fig.~\ref{fig:quest_rpss}c).

\aifssubs{}* scores slightly below \aifsgaia{}, which we attribute to its
fine-tuning on operational analyses: initialised here from ERA5, the model must
accommodate a distribution shift in its initial conditions and loses some skill.
The same effect applies to \aifsens{}*. Its higher spatial resolution
($\sim$0.25\textdegree) and 6\,h time step additionally lead to strong biases
under autoregressive rollout beyond the medium range. We include it to
illustrate the value of two design priorities: adapting models specifically for
longer timescales, and ensuring they generalise across different initial conditions.

The spatial distribution of week\,3 skill is shown in
Fig.~\ref{fig:quest_maps}. The absolute RPSS for \aifsgaia{}
(Fig.~\ref{fig:quest_maps}a, c, e) is positive (green) across most of the globe
for all three variables, with mean sea-level pressure exhibiting the most
spatially coherent skill, while regions of negative skill (brown) are
concentrated in the deep tropics for 2\,m temperature and over scattered land
regions for total precipitation. Differences in RPSS between \aifsgaia{} and
the IFS (Fig.~\ref{fig:quest_maps}b, d, f) are noisy on this short sample, but
several patterns emerge from the bootstrap. For 2\,m temperature and mean
sea-level pressure, \aifsgaia{} shows statistically significant improvements
over Europe and Eurasia, accompanied by a significant degradation of skill
over the Tropical Pacific. For total precipitation, significant improvements
appear in the tropics, particularly over Asia; we caution that several of
these regions have very low climatological precipitation, which can inflate
relative skill scores.

The six-month evaluation window for the real-time AI Weather Quest remains
short, and the results should be interpreted with this sample size in mind.

\section{Discussion}
\label{sec:conclusions}

We have adapted the \aifscrps{} medium-range model for sub-seasonal prediction.
\aifssubs{} extends the autoregressive time step to 24\,h to limit error
accumulation, adds stratospheric levels and top-of-atmosphere thermal radiation
as predictors, and reserves 2007--2011 for independent verification. Unlike the
medium-range setting, which optimises real-time forecasts only, \aifssubs{}
targets calibrated anomalies defined relative to its reforecast climatology. 
Our mixed fine-tuning protocol allows generalisation across ERA5 and operational
initial conditions.

Across weeks\,2--6 \aifssubs{} matches the operational IFS in probabilistic skill
while significantly reducing systematic biases. For the convective (OLR)
component of the MJO, \aifssubs{} extends skilful forecasts by approximately
eight days, while matching or exceeding the IFS for the full multivariate RMM
index. In addition, \aifssubs{} reproduces sudden stratospheric warming frequency
and surface impact comparable to the IFS. It also captures the observed MJO
modulation of tropical cyclone activity as well as the IFS, lending confidence
to its sub-seasonal forecasts of tropical cyclone activity. In real-time
submissions to the AI Weather Quest, \aifsgaia{} shows the highest
variable-averaged ranked probability skill score at weeks\,3 and 4 among fully
ML models and is narrowly ahead of the IFS. 

Several limitations point to future improvements. \aifssubs{} forecasts the
atmosphere alone. At sub-seasonal lead times, however, much of the
predictability arises from the slowly varying boundary conditions such as the
land, surface ocean and sea ice. Coupling \aifssubs{} to land and ocean
components is therefore a natural next step \citep{hahner2026}, and one we
expect to be particularly beneficial for surface variables over land and in the
tropics.

The 24\,h time step limits error accumulation over the long autoregressive
rollout, but it comes at a cost: the model sees only instantaneous fields at a
single time of day and cannot resolve the diurnal cycle. In future work we aim
to stabilise error accumulation so that we can return to a 6\,h time step, or
alternatively predict daily means rather than instantaneous snapshots, either of
which would recover sub-daily information without sacrificing rollout stability.

\aifssubs{} is trained on ERA5 and fine-tuned on operational analyses.
Incorporating additional data sources, in particular operational reforecasts,
could expand the training data and could allow the model to
identify predictable patterns from non-predictable ones.

Finally, the low inference cost of \aifssubs{} --- about 200 times less energy
per forecast than the IFS (Sec.~\ref{sec:model}) --- opens a more immediate
opportunity: real-time ensembles of the order of 1000 members that could yield
more accurate probabilistic sub-seasonal forecasts.

\section*{Acknowledgements}

We thank the whole AIFS team at ECMWF, and in particular Cathal
O'Brien for his support in running these models on the EuroHPC supercomputers
and in estimating their compute cost. 

We acknowledge the EuroHPC Joint Undertaking for awarding us access to the
EuroHPC supercomputers MN5, hosted by BSC in Barcelona, and JUPITER, hosted by
the Jülich Supercomputing Centre.


\printbibliography

\clearpage
\appendix
\renewcommand{\thefigure}{\thesection.\arabic{figure}}
\counterwithin{figure}{section}

\section{Supplementary Material}

\subsection{Evaluation protocol}
  \label{sec:si_evaluation}

\subsubsection{Realtime forecast evaluation}
  \label{sec:si_realtime}
For each forecast target date $t$ and lead time $\tau$, we construct the model
climatology from a 10-member ensemble of hindcasts initialised on the five days
$\{-4, -2, 0, +2, +4\}$ around the corresponding month--day in each of the 20
preceding years. This yields $N_{\mathrm{c}} = 5 \times 20 \times 10 = 1000$
hindcast samples per target date, from which the climatological mean and
distribution are estimated. Forecast anomalies at grid cell $g$ are defined as

\begin{equation}
  \hat{x}'_i(\tau, g) \;=\; \hat{x}_i(\tau, g) \;-\; \overline{\hat{x}}_{\mathrm{c}}(\tau, g),
  \qquad
  \overline{\hat{x}}_{\mathrm{c}}(\tau, g) \;=\; \frac{1}{N_{\mathrm{c}}}
  \sum_{j=1}^{N_{\mathrm{c}}} \hat{x}^{\mathrm{c}}_j(\tau, g),
  \label{eq:anomaly}
\end{equation}

where $\hat{x}^{\mathrm{c}}_j$ denotes the $j$-th hindcast sample. ERA5 serves
as the ground truth; its climatology $\overline{x}_{\mathrm{c}}(\tau, g)$ is
built analogously from a $[-4, +4]$\,day window over the same 20-year period,
and observed anomalies are defined as $x'(\tau, g) = x(\tau, g) -
\overline{x}_{\mathrm{c}}(\tau, g)$. This ensures that forecast and observed
anomalies share a consistent bias-corrected reference.

Following the AI Weather Quest protocol, we bin the climatological distribution
into five equiprobable categories, bounded by the quintile thresholds
$q_{k/5}(\tau, g)$ for $k = 1, \ldots, 4$, with $q_0 = -\infty$ and
$q_{5/5} = +\infty$. The forecast probability that the verification falls in
bin $k$ is

\begin{equation}
  p_k(\tau, g) \;=\; \frac{1}{N_{\mathrm{e}}} \sum_{i=1}^{N_{\mathrm{e}}}
  \mathbf{1}\!\left[\, q_{(k-1)/5}(\tau, g) \;<\; \hat{x}_i(\tau, g) \;\le\;
  q_{k/5}(\tau, g) \,\right],
  \label{eq:quintile_prob}
\end{equation}

where $N_{\mathrm{e}}$ is the number of forecast ensemble members and
$\mathbf{1}[\cdot]$ the indicator function.

\subsubsection{Mean absolute bias score}
\label{sec:si_mabs}

For each variable and lead time $\tau$, the systematic bias at grid cell $g$ is
the mean error of the ensemble-mean forecast over the $N_{\mathrm{f}}$ forecast
cases,
\begin{equation}
  b(\tau, g) \;=\; \frac{1}{N_{\mathrm{f}}} \sum_{n=1}^{N_{\mathrm{f}}}
  \left( \overline{\hat{x}}_n(\tau, g) - x_n(\tau, g) \right),
  \label{eq:bias}
\end{equation}
where $\overline{\hat{x}}_n$ is the ensemble mean and $x_n$ the ERA5
verification. The mean absolute bias (MAB) is the area-weighted average of
$\lvert b \rvert$ over a region $\mathcal{R}$,
\begin{equation}
  \mathrm{MAB}(\tau) \;=\; \sum_{g \in \mathcal{R}} w_g \, \lvert b(\tau, g) \rvert,
  \qquad w_g \propto \cos\phi_g, \quad \sum_{g \in \mathcal{R}} w_g = 1,
  \label{eq:mab}
\end{equation}
with latitude $\phi_g$. The mean absolute bias score (MABS) is the relative
change in MAB against the reference,
\begin{equation}
  \mathrm{MABS}(\tau) \;=\; 1 - \frac{\mathrm{MAB}_{\mathrm{model}}(\tau)}{\mathrm{MAB}_{\mathrm{ref}}(\tau)},
  \label{eq:mabs}
\end{equation}
so that $\mathrm{MABS} > 0$ indicates a smaller bias than the reference.

\subsubsection{Fair continuous ranked probability score}
\label{sec:si_fcrps}

For an $N_{\mathrm{e}}$-member anomaly forecast $\hat{X}' = \{\hat{x}'_i\}$
verifying against $x'$, the fair CRPS
\citep{ferro2008,ferro2014,leutbecher2019} is
\begin{equation}
  \mathrm{fCRPS}(\hat{X}', x') \;=\; \frac{1}{N_{\mathrm{e}}} \sum_{i=1}^{N_{\mathrm{e}}} \lvert \hat{x}'_i - x' \rvert
  \;-\; \frac{1}{2 N_{\mathrm{e}}(N_{\mathrm{e}}-1)} \sum_{i=1}^{N_{\mathrm{e}}} \sum_{j=1}^{N_{\mathrm{e}}} \lvert \hat{x}'_i - \hat{x}'_j \rvert .
  \label{eq:fcrps}
\end{equation}
The $(N_{\mathrm{e}}-1)$ normalisation removes the finite-ensemble bias of the
standard estimator, so the score is unbiased with respect to ensemble size. The
area-weighted regional score at lead $\tau$ averages over grid cells and
forecast cases,
\begin{equation}
  \overline{\mathrm{fCRPS}}(\tau) \;=\; \frac{1}{N_{\mathrm{f}}} \sum_{n=1}^{N_{\mathrm{f}}}
  \sum_{g \in \mathcal{R}} w_g \, \mathrm{fCRPS}\!\left( \hat{X}'_n(\tau, g), x'_n(\tau, g) \right),
  \label{eq:fcrps_regional}
\end{equation}
and the corresponding skill score against the reference is
\begin{equation}
  \mathrm{fCRPSS}(\tau) \;=\; 1 - \frac{\overline{\mathrm{fCRPS}}_{\mathrm{model}}(\tau)}{\overline{\mathrm{fCRPS}}_{\mathrm{ref}}(\tau)} .
  \label{eq:fcrpss}
\end{equation}

\subsection{MJO forecasts}

\begin{figure}[H]
  \centering
  \includegraphics[width=\linewidth]{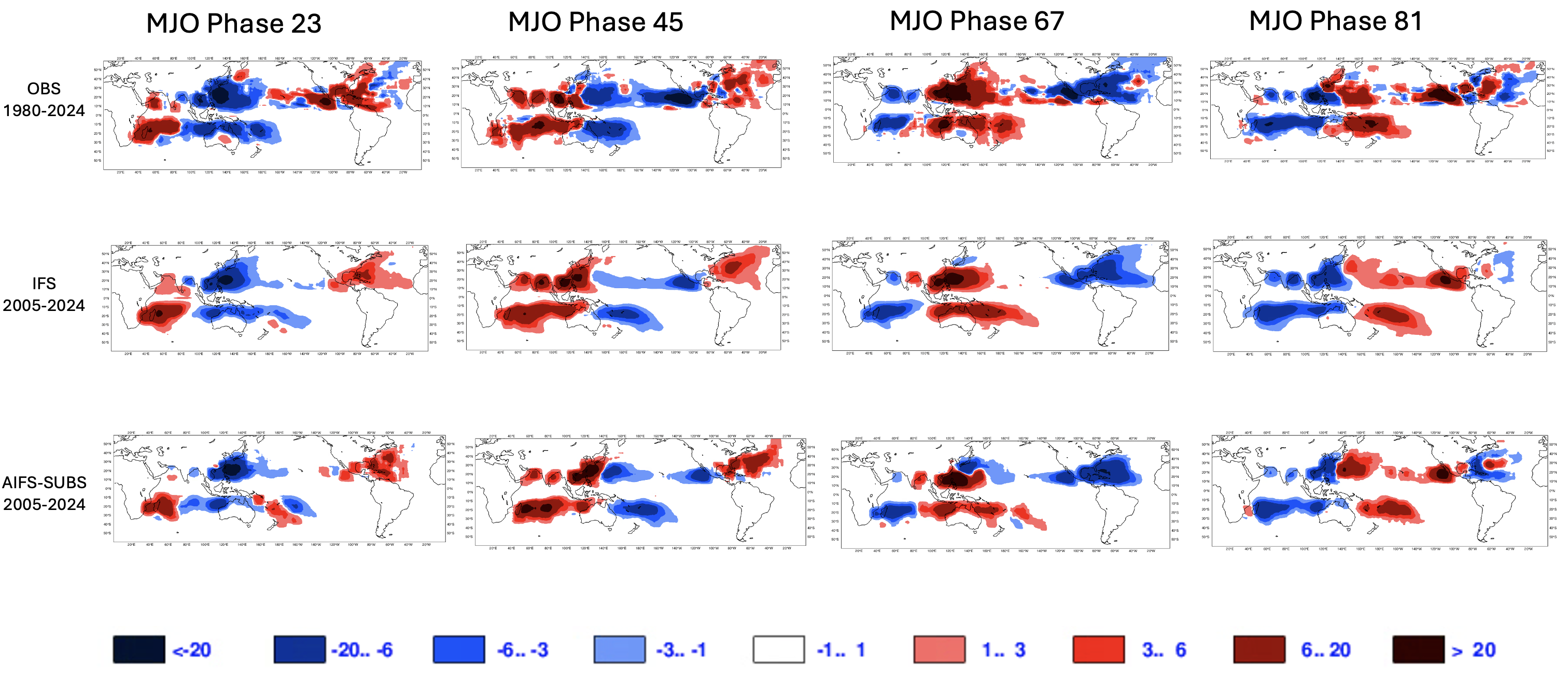}
  \caption{Same as Fig.~\ref{fig:mjo_tc_teleconnection} but for years within the
  training period, i.e. 2005-2024, excluding the verification period 2007-2011.
  The anomaly number of TC days per MJO phase shows a similar pattern as in the
  out-of-training verification period.}
  \label{fig:si_mjo_tc_teleconnection}
\end{figure}

\subsection{Surface impact of SSW events in the IFS}

\begin{figure}[H]
  \centering
  \includegraphics[width=\linewidth]{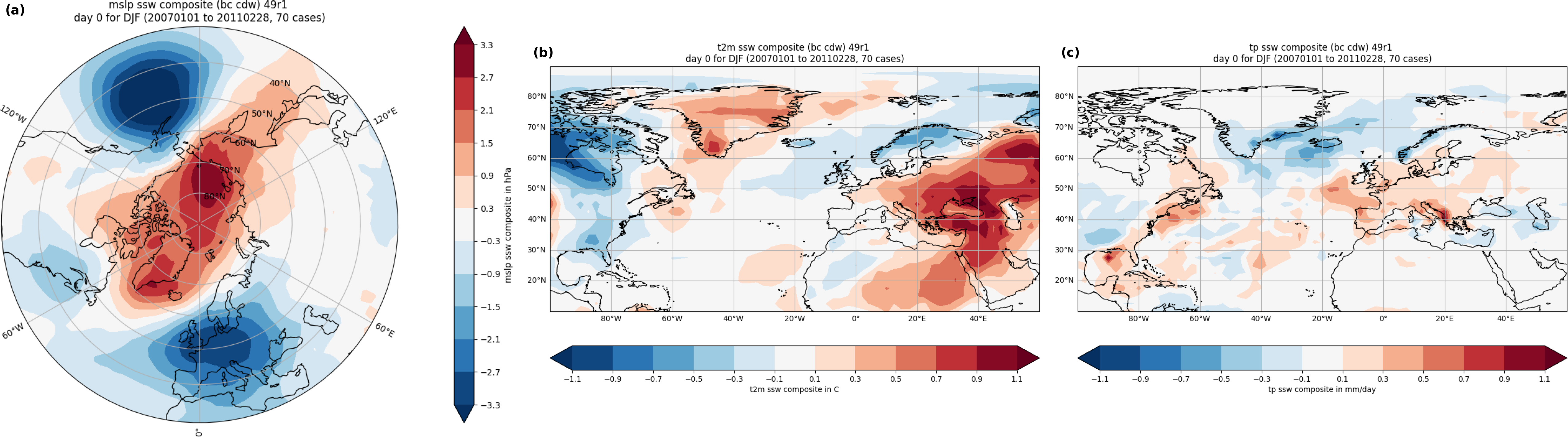}
  \caption{Same as Fig.~\ref{fig:ssw} but for the 49R1 IFS reforecasts. Surface impact
  composites 10-35 days from all SSW events occurring in the 2007-2011
  reforecasts for (a) mean sea level pressure, (b) 2m temperature, and (c) total
  precipitation.}
  \label{fig:si_ssw_ifs} 
\end{figure}

\end{document}